\documentclass[times, twoside]{zHenriquesLab-StyleBioRxiv}
\usepackage{blindtext}
\usepackage{booktabs}
\usepackage{multirow}

\leadauthor{Giedziun} 

\begin{document}

% insert title
\title{RF-DETR for Robust Mitotic Figure Detection: A MIDOG 2025 Track 1 Approach}

% insert title for footer
\shorttitle{RF-DETR for MIDOG 2025}

% Use letters for affiliations, numbers to show equal authorship (if applicable) and to indicate the corresponding author
\author[1,2]{Piotr Giedziun}
\author[2]{Jan Sołtysik}
\author[2]{Mateusz Górczany}
\author[1,2]{Norbert Ropiak}
\author[1,2]{Marcin Przymus}
\author[2]{Piotr Krajewski}
\author[2]{Jarosław Kwiecień}
\author[3,4]{Artur Bartczak}
\author[5]{Izabela Wasiak}
\author[6,7,8]{Mateusz Maniewski}
\affil[1]{Wrocław University of Science and Technology, Wroclaw, Poland}
\affil[2]{Cancer Center Sp. z o. o., Wroclaw, Poland}
\affil[3]{Hospital for Lung Diseases - Rebirth, Zakopane, Poland}
\affil[4]{Maria Sklodowska-Curie National Research Institute of Oncology, Krakow, Poland}
\affil[5]{Pathology Department, 10th Military Research Hospital in Bydgoszcz, Poland}
\affil[6]{Department of Tumor Pathology and Pathomorphology, Oncology Centre Prof. Franciszek Łukaszczyk Memorial Hospital, Bydgoszcz, Poland}
\affil[7]{Doctoral School of Medical and Health Sciences, Nicolaus Copernicus University in Toruń, Bydgoszcz, Poland}
\affil[8]{Department of Obstetrics, Gynaecology and Oncology, Collegium Medicum in Bydgoszcz, Nicolaus Copernicus University in Torun, Poland}

\maketitle

%TC:break Abstract
\begin{abstract}
	Mitotic figure detection in histopathology images remains challenging due to significant domain shifts across different scanners, staining protocols, and tissue types. This paper presents our approach for the MIDOG 2025 challenge Track 1, focusing on robust mitotic figure detection across diverse histological contexts. While we initially planned a two-stage approach combining high-recall detection with subsequent classification refinement, time constraints led us to focus on optimizing a single-stage detection pipeline. We employed RF-DETR (Roboflow Detection Transformer) with hard negative mining, trained on MIDOG++ dataset. On the preliminary test set, our method achieved an F1 score of 0.789 with a recall of 0.839 and precision of 0.746, demonstrating effective generalization across unseen domains. The proposed solution offers insights into the importance of training data balance and hard negative mining for addressing domain shift challenges in mitotic figure detection.
\end{abstract}
%TC:break main


\begin{corrauthor}
	piotr.giedziun@cancercenter.ai
\end{corrauthor}

\section*{Introduction}

Mitotic figure (MF) detection is a crucial component in cancer grading and prognosis assessment, reflecting tumor proliferation. In breast cancer specifically, this metric is central to grading \cite{vanDiest675}. Yet substantial inter- and intra-observer discrepancies persist—both in selecting the mitotic count region of interest (MC-ROI) and in distinguishing MFs from mitotic-like figures (apoptotic cells, compressed nuclei, cellular debris) - and are compounded by variability across labs, scanners, and stains. Computer assistance may help reduce this variability and improve standardization and MF detection \cite{03009858211067478}.

The MIDOG challenge series has emerged as a benchmark for evaluating the robustness of mitotic figure detection algorithms under domain shift conditions. The 2025 edition introduces additional complexities by expanding the scope beyond traditional tumor hotspot regions to include challenging areas such as necrotic zones, inflammatory regions, and non-tumor tissue \cite{ammeling_mitosis_2025}.

Previous MIDOG challenges have demonstrated the difficulty of this task, with top-performing algorithms in MIDOG 2022 achieving top F1 score of 0.764 on unseen domains. These results highlight both the progress made in domain generalization and the remaining gap toward clinical applicability. Recent approaches have explored various strategies including domain adversarial training, extensive data augmentation, and ensemble methods to improve cross-domain performance \cite{AUBREVILLE2024103155}.

In this work, we present our approach to the MIDOG 2025 challenge Track 1. Our initial strategy involved developing a two-stage pipeline: first, a high-recall detection stage to identify all potential mitotic figure candidates, followed by a specialized classification stage to filter false positives and improve precision. This approach was motivated by the observation that detection and fine-grained classification might benefit from different architectural designs and training strategies. However, due to time constraints in the challenge timeline, we focused our efforts on optimizing the first stage – the detection component - which ultimately served as our complete solution.

Our final approach leverages RF-DETR (Roboflow Detection Transformer), a transformer-based detection architecture that employs deformable attention mechanisms and a DINOv2 backbone. This architecture was selected for its demonstrated strength in detecting ambiguous objects with high morphological similarity to background elements. \cite{sapkota2025rfdetrobjectdetectionvs}.

\section*{Datasets}

Our training pipeline was meant to utilize multiple publicly available mitotic figure detection datasets to ensure broad domain coverage:

\textbf{MIDOG++ Dataset:} The primary training resource, containing 11,937 mitotic figure annotations across 503 cases from seven tumor types (nine domain variants) \cite{midogpp2023}. This dataset encompasses various tumor types including human breast carcinoma, canine lung adenocarcinoma, and canine tumors, captured using different scanners and staining protocols. The diversity of this dataset forms the foundation for domain generalization.

\textbf{MITOS\_WSI\_CCMCT Dataset:} A comprehensive collection of 32 fully annotated whole-slide images of canine cutaneous mast cell tumors, containing over 40,000 mitotic figure annotations \cite{aubreville_completely_2020}. This dataset is particularly valuable for its exhaustive annotations across entire slides, including non-hotspot regions that contain challenging morphological patterns.

\textbf{MITOS\_WSI\_CMC Dataset:} Comprising 21 fully annotated whole-slide images of canine mammary carcinoma \cite{aubreville_completely_2020}, this dataset provides additional domain variety in tissue morphology and staining characteristics.

\textbf{SPIDER Dataset:} We utilized SPIDER-Skin and SPIDER-Breast subsets from the Multi-Organ Supervised Pathology Dataset \cite{nechaev2025spidercomprehensivemultiorgansupervised} to extract necrosis regions for hard negative mining. We extracted 5,000 regions of 360×360 pixels (center crop) from annotated necrotic areas to boost true negative detection in challenging tissue regions.

We have processed these datasets and created train, valid, and test splits (0.7, 0.15, 0.15) with 380×380 px patches.

\section*{Methods}

\subsection*{Model Architecture}

We used RF-DETR as the base detection architecture. Attempts to adapt it for small, rare objects - by varying the numbers of samples and queries - did not improve metrics. In a patch-size ablation, 380×380 patches performed best; using 896×896 patches decreased the F1 score.

\subsection*{Training Strategy}

Training was performed using the AdamW optimizer with a cosine learning rate schedule. We employed standard data augmentation techniques including: random horizontal and vertical flips, random resize to 400, 500, 600, and random size crop to 384.

The model was trained for 50 epochs with early stopping based on validation performance on mAP\@50. We used batch sizes ranging from 4 to 32, with gradient accumulation steps inversely scaled (8 steps for batch size 4, down to 2 steps for batch size 32) to maintain an effective batch size of 32-64 while accommodating GPU memory constraints.

\subsection*{Best performing model}

We trained RF-DETR variants (Nano, Small, Medium, Base, Large) on various subsets and combinations of the datasets described above.
Our top-performing model was RF-DETR–Large with an exponential moving average (EMA) of the weights (decay of 0.993), trained exclusively on MIDOG++.

Notably, the same architecture and hyperparameters applied to different training/validation splits of MIDOG++ produced lower scores, underscoring how the data/domain distribution across training, validation, and test sets critically influences performance.

Our second-best submission to the MIDOG 2025 challenge was RF-DETR–Base with EMA, trained only on the MIDOG++ breast-cancer subset.

\section*{Results}

\subsection*{Performance on Preliminary Test Set}

Our best model - RF-DETR Large was evaluated on the MIDOG 2025 preliminary test set, providing a rigorous assessment of domain generalization capabilities. The test set incorporates both traditional regions of interest and challenging areas including necrosis and inflammation.

\textbf{Overall Performance Metrics:}
\begin{itemize}
	\item F1 Score: 0.789
	\item Recall: 0.839
	\item Precision: 0.746
\end{itemize}

The high recall of 0.839 indicates that our model successfully identifies the majority of true mitotic figures, while the precision of 0.746 suggests reasonable discrimination against false positives. The F1 score of 0.789 represents a balanced performance between these two metrics.

\subsection*{Per-Domain Analysis}

The model demonstrated consistent performance across different tumor types in the preliminary test set. The relatively consistent F1 scores across tumor types (ranging from 0.767 to 0.810) suggest effective domain generalization. Notably, Tumor Type 2 showed the highest recall (0.896) but lower precision (0.671), indicating potential differences in tissue characteristics that affect the false positive rate.

\subsection*{Comparison with Initial Two-Stage Design}

While we were unable to complete the planned two-stage approach within the challenge timeline, our single-stage RF-DETR solution achieved competitive results. The high recall (0.839) suggests that the detection stage we optimized would have provided a strong foundation for the planned classification refinement stage. The precision of 0.746, while respectable, indicates potential room for improvement that the second stage might have addressed.

\section*{Discussion}

Our results demonstrate that state-of-the-art detection models are able to achieve competitive performance in the mitosis detection problem.

\subsection*{Key Findings}

\paragraph{Domain Balance is Crucial:} Exposure to different tissue types during training is vital for model performance. However, the score of breast-only RF-DETR Base may suggest that exposure to different tissue types could be more important than to a specific cancer type.

\paragraph{RF-DETR Architecture:} The transformer-based detection architecture demonstrated strong capability for small object detection in histopathology images.

\paragraph{Normalization and Augmentation Techniques:} We evaluated multiple data normalization methods (Macenko, Multi-Macenko, Reinhard, Hematoxylin-only, Eosin-only) and augmentation techniques (CutMix, Gaussian blur). All of these approaches yielded lower F1 scores compared to training on the unmodified dataset.

\section*{Conclusion}

We presented a single-stage approach for mitotic figure detection in the MIDOG 2025 challenge, achieving an F1 score of 0.789 on the preliminary test set. Fine-tuned RF-DETR detector with domain-balanced training and hard negative mining demonstrated strong cross-domain generalization. The consistent performance across different tumor types validates our training strategy's effectiveness.

Future work should explore the potential benefits of the two-stage approach, combining high-recall detection with specialized classification for false positive reduction. Furthermore, incorporating explicit domain adaptation techniques and test-time strategies could further improve robustness. As the field progresses toward clinical deployment, approaches that balance accuracy, generalization, and computational efficiency will be essential for practical adoption in digital pathology workflows.

\begin{acknowledgements}
We thank the organizers of the MIDOG 2025 challenge for providing this valuable benchmark and the computational pathology community for making datasets publicly available.
\end{acknowledgements}

\section*{Bibliography}
\bibliography{literature}

\end{document}